\documentclass[aps,prl,twocolumn,floatfix,nofootinbib,showpacs]{revtex4}
\usepackage{graphicx}
\usepackage{color}

\def\beq{\begin{equation}}
\def\eeq{\end{equation}}
\def\eeqn{\end{equation}}
\newcommand\iden{\leavevmode\hbox{\small1\normalsize\kern-.33em1}}

\newcommand{\bea} {\begin{eqnarray}}
\newcommand{\eea} {\end{eqnarray}}

\let\jnfont=\rm
\def\NPB#1,{{\jnfont Nucl.\ Phys.\ B }{\bf #1},}
\def\PLB#1,{{\jnfont Phys.\ Lett.\ B }{\bf #1},}
\def\EPJC#1,{{\jnfont Eur.\ Phys.\ Jour.\ C }{\bf #1},}
\def\PRD#1,{{\jnfont Phys.\ Rev.\ D }{\bf #1},}
\def\PRL#1,{{\jnfont Phys.\ Rev.\ Lett.\ }{\bf #1},}
\def\MPLA#1,{{\jnfont Mod.\ Phys.\ Lett.\ A }{\bf #1},}
\def\JPG#1,{{\jnfont J.\ Phys.\ G }{\bf #1},}
\def\CTP#1,{{\jnfont Commun.\ Theor.\ Phys.\ }{\bf #1},}
\def\JHEP#1,{{\jnfont JHEP \ }{\bf #1},}
\def\NPPS#1,{{\jnfont Nucl.\ Phys.\ Proc.\ Suppl.\ }{\bf #1},}
\def\CPC#1,{{\jnfont Computl.\ Phys.\ Commun.\ }{\bf #1},}
\def\CPL#1,{{\jnfont Chin.\ Phys.\ Lett. }{\bf #1},}
\def\AJS#1,{{\jnfont Astrophys.\ J.\ Suppl. }{\bf #1},}
\def\PR#1,{{\jnfont Phys.\ Rept. }{\bf #1},}
\def\AP#1,{{\jnfont Astropart.\ Phys. }{\bf #1},}
\def\EPL#1,{{\jnfont Europhys.\ Lett. }{\bf #1},}
\def\FP#1,{{\jnfont Fortsch.\ Phys. }{\bf #1},}

\begin{document}

\title{The simple interpretations of lepton anomalies in the
lepton-specific inert two-Higgs-doublet model}
\renewcommand{\thefootnote}{\fnsymbol{footnote}}

\author{Xiao-Fang Han$^{1}$, Tianjun Li$^{2,3}$, Lei Wang$^{1}$, Yang Zhang$^{4}$}
 \affiliation{$^1$ Department of Physics, Yantai University, Yantai
264005, P. R. China\\
$^2$ CAS Key Laboratory of Theoretical Physics, Institute of Theoretical Physics,
 Chinese Academy of Sciences, Beijing 100190,  P. R. China   \\
$^3$ School of Physical Sciences, University of Chinese Academy of Sciences, Beijing 100049,  P. R.  China\\
$^4$ ARC Centre of Excellence for Particle Physics at the Tera-scale, School of Physics and Astronomy, Monash University, Melbourne, Victoria 3800, Australia}
\renewcommand{\thefootnote}{\arabic{footnote}}

\date{\today}

\begin{abstract}
There exist about $3.7\sigma$ positive and $2.4\sigma$ negative deviations
 in the muon and electron anomalous magnetic moments ($g-2$). Also,
some ratios for lepton universality in $\tau$ decays have 
almost $2\sigma$ deviations from the Standard Model. In this paper,
 we propose a lepton-specific inert two-Higgs-doublet model. 
After imposing all the relevant theoretical and experimental constraints, 
we show that these lepton anomalies can be explained simultaneously
in many parameter spaces with $m_H > 200$ GeV and $m_A~(m_{H^\pm})> 500$ GeV 
for appropriate Yukawa couplings between leptons and inert Higgs. The key point 
is that these Yukawa couplings for $\mu$  and  $\tau$/$e$ have opposite sign. 
\end{abstract}
\maketitle

{\bf Introduction~--}~ 
The Standard Model (SM) describes the elementary particles,
as well as the fundamental interactions between them.
In particular, such description is sensitive to the quantum corrections.
For example, since Schwinger's seminar calculation of the electron 
anomalous magnetic moment $a_{e}=\alpha/2\pi$ \cite{winger},  the  charged lepton
anomalous magnetic moments have become the powerful precision tests of
Quantum Electrodynamics (QED), and subsequently the full SM.
The muon anomalous magnetic moment $g-2$ has been
a long-standing puzzle since the announcement by the E821 experiment
in 2001~\cite{mug2-exp}. The experimental value has an approximate $3.7\sigma$
discrepancy from the SM prediction \cite{mug2-3.7}
\bea
\Delta a_\mu=a_\mu^{exp}-a_\mu^{SM}=(274\pm73)\times10^{-11}.
\eea 

Very recently, an improvement in the measured mass of atomic Cesium
used in conjunction with other known mass ratios and
the Rydberg constant leads to the most precise value of the fine structure
constant \cite{alpha-exp}. As a result, the experimental value of the electron $g-2$ has a
$2.4\sigma$ deviation from the SM prediction \cite{eg2-theory,eg2-2.4-1,eg2-2.4-2}
\bea
\Delta a_e=a_e^{exp}-a_e^{SM}=(-87\pm36)\times10^{-14},
\eea
which has opposite in sign from the muon $g-2$.

The Lepton Flavor Universality (LFU) in the $\tau$ decays is an excellent way
to probe new physics.
The HFAG Collaboration reported three ratios from pure leptonic processes, and two ratios
from semi-hadronic processes, $\tau \to \pi/K \nu$ and $\pi/K \to \mu \nu$ \cite{tauexp}
\begin{eqnarray} \label{hfag-data}
&&
\left( g_\tau \over g_\mu \right) =1.0011 \pm 0.0015,~~
\left( g_\tau \over g_e \right) = 1.0029 \pm 0.0015,~~ \nonumber\\
&&
\left( g_\mu \over g_e \right) = 1.0018 \pm 0.0014, 
\left( g_\tau \over g_\mu \right)_\pi = 0.9963 \pm 0.0027,\nonumber\\
&&
\left( g_\tau \over g_\mu \right)_K = 0.9858 \pm 0.0071,
\end{eqnarray}
 where the ratios of $g_\tau/g_e$ and $\left( g_\tau \over g_\mu \right)_K$ 
have almost $2\sigma$ deviations from the SM.

Muon $g-2$ anomaly can be simply explained in the lepton-specific 
two-Higgs-doublet model (2HDM) and 
aligned 2HDM. However, the tree-level diagram mediated by the charged Higgs gives negative contribution
to the decay $\tau\to \mu\nu\bar{\nu}$ \cite{tavv-1,crivellin,tavv-2}, which will raise the discrepancy in the LFU in
$\tau$ decays. In addition, these two types of 2HDM do not explain the muon and electron $g-2$
 simultaneously since there is an opposite sign between them.
Therefore, we shall propose a lepton-specific inert 2HDM 
to explain all three anomalies of muon and electron $g-2$ as well as  
LFU in $\tau$ decay simultaneously. 
In our model, for the extra Higgses ($H,A,H^{\pm}$), the Yukawa couplings for $\mu$ and $\tau$/$e$ have opposite sign.
In 2012, G. F. Giudice et al. used the approach of the effective operator to discuss the contributions of light scalar to
the muon and electron $g-2$. The contributions of two-loop Barr-Zee type diagrams can be positive or negative depending
on the relative sign of the Yukawa couplings for muon, electron, and tau \cite{12086583}.
Although the muon and electron $g-2$ have been addressed
simultaneously in a few recent papers~\cite{mueg2-1,mueg2-2,mueg2-3,mueg2-4,mueg2-5},  
it seems to us that our model is simpler from the renormalized theory point of view.

{\bf Lepton-specific inert 2HDM~--}~We introduce an inert Higgs doublet $\Phi_2$ in the SM
 as well as a discrete $Z_2$ symmetry under which $\Phi_2$ is odd while 
all the SM particles are even.
The scalar potential for the SM Higgs field $\Phi_1$ and inert doublet $\Phi_2$  is
\begin{eqnarray} \label{V2HDM} \mathrm{V} &=& Y_1
(\Phi_1^{\dagger} \Phi_1) + Y_2 (\Phi_2^{\dagger}
\Phi_2)+ \frac{\lambda_1}{2}  (\Phi_1^{\dagger} \Phi_1)^2 +
\frac{\lambda_2}{2} (\Phi_2^{\dagger} \Phi_2)^2  \nonumber \\
&&+ \lambda_3
(\Phi_1^{\dagger} \Phi_1)(\Phi_2^{\dagger} \Phi_2) + \lambda_4
(\Phi_1^{\dagger}
\Phi_2)(\Phi_2^{\dagger} \Phi_1)\nonumber \\
&&+ \left[\frac{\lambda_5}{2} (\Phi_1^{\dagger} \Phi_2)^2 + \rm
h.c.\right]~.
\end{eqnarray}
We focus on the CP-conserving case where all
$\lambda_i$ are real. The two complex
scalar doublets can be written as
\begin{equation} \label{field}
\Phi_1=\left(\begin{array}{c} G^+ \\
\frac{1}{\sqrt{2}}\,(v+h+iG_0)
\end{array}\right)\,, \ \ \
\Phi_2=\left(\begin{array}{c} H^+ \\
\frac{1}{\sqrt{2}}\,(H+iA)
\end{array}\right). \nonumber
\end{equation}
The $\Phi_1$ field has the vacuum expectation value (VEV) $v=$246
GeV, and the VEV of $\Phi_2$ field is zero. $Y_1$ is fixed by the scalar
potential minimization condition.
The $H^+$ and $A$ are the mass eigenstates of the charged Higgs boson and
CP-odd Higgs boson.  
 Their masses are given as
\beq \label{masshp}
 m_{H^\pm}^2  = Y_2+\frac{\lambda_3}{2} v^2, ~~m_{A}^2  = m_{H^\pm}^2+\frac{1}{2}(\lambda_4-\lambda_5) v^2.
 \eeq
The $h$ and $H$ have no mixing, and they are two mass eigenstates of the CP-even Higgses.
In this paper, we take the light CP-even Higgs $h$ as the SM-like Higgs. Their masses are given as
\beq \label{massh}
 m_{h}^2  = \lambda_1 v^2\equiv (125~{\rm GeV })^2, ~~m_{H}^2  = m_{A}^2+\lambda_5 v^2.
 \eeq

The fermions obtain the mass terms from the Yukawa interactions with $\Phi_1$
 \beq \label{yukawacoupling} - {\cal L} = y_u\overline{Q}_L \,
\tilde{{ \Phi}}_1 \,u_R + y_d\overline{Q}_L\,{\Phi}_1 \, d_R +  y_l\overline{L}_L \, {\Phi}_1
\, e_R + \mbox{h.c.}, \eeq
where $Q_L^T=(u_L\,,d_L)$, $L_L^T=(\nu_L\,,l_L)$, 
$\widetilde\Phi_{1}=i\tau_2 \Phi_{1}^*$, and $y_u$, $y_d$ and $y_\ell$ are
$3 \times 3$ matrices in family space. In addition, only in the lepton sector we introduce 
the $Z_2$ symmetry-breaking Yukawa interactions of $\Phi_2$,
 \bea \label{phi2coupling} - {\cal L} &=&  \sqrt{2}~\kappa_e \,\overline{L}_{1L} \, {\Phi}_2
\, e_R  \, + \sqrt{2}~\kappa_\mu\, \overline{L}_{2L} \, {\Phi}_2
\,\mu_R \, \nonumber\\
&&+\, \sqrt{2}~\kappa_\tau \,\overline{L}_{3L} \, {\Phi}_2
\,\tau_R\, + \, \mbox{h.c.}\,. \eea
Such the $Z_2$ symmetry-breaking effect only for the lepton sector 
can be realized in the high-dimensional brane world scenario, which will be studied elsewhere.
From Eq. (\ref{phi2coupling}), we can obtain the lepton Yukawa couplings of extra Higgses ($H$, $A$, and $H^\pm$).
The neutral Higgses $A$ and $H$ have no couplings to $ZZ, ~WW$.

{\bf Numerical results~--}~According to Eqs. (\ref{masshp}) and (\ref{massh}), 
the values of $\lambda_1$, $\lambda_5$ and $\lambda_4$ can
be determined by $m_h$ ($=125~{\rm GeV}$), $m_H$, $m_A$ and $m_{H^\pm}$.
$\lambda_2$ controls the quartic couplings of extra Higgses, but does not affect 
the physics observables. So we simply take $\lambda_2=\lambda_1$. 
Because the precision electroweak data favor small mass splitting between $m_A$ and $m_{H^\pm}$, 
we simply choose $m_A=m_{H^\pm}$. We employ the $\textsf{2HDMC}$ \cite{2hc-1}
to implement the theoretical constraints from vacuum stability, unitarity and
perturbativity, as well as the constraints of
the oblique parameters ($S$, $T$, $U$).
We scan over several key parameters in the following ranges
\bea
&&0.5 <\kappa_\tau<1,~~-0.25<\kappa_\mu<0,~~0<\kappa_e<0.01,\nonumber\\
&&200~{\rm GeV}<m_H<350~{\rm GeV},\nonumber\\ &&500~{\rm GeV} <m_A=m_{H^\pm}<700~{\rm GeV}.
\eea
In such ranges of $\kappa_\tau$, $\kappa_\mu$ and $\kappa_e$, the corresponding Yukawa couplings do not become non-perturbative.
At the tree-level, the SM-like Higgs has the same couplings to the SM particles as the SM, and no exotic decay mode.
The masses of extra Higgses are beyond the exclusion range of the searches for the neutral and charged Higgs at the LEP.
Since the extra Higgses have no couplings to quarks due to $Z_2$ symmetry, 
we can safely neglect the limits from the observables of meson.
The extra Higgs bosons are dominantly produced at the LHC via electroweak processes.
We generate the Monte Carlo events using \texttt{MG5\_aMC-2.4.3}~\cite{Alwall:2014hca} with \texttt{PYTHIA6}~\cite{Torrielli:2010aw}, and adopt the constraints from all the analysis for the 13 TeV LHC in \texttt{CheckMATE 2.0.7}~\cite{Dercks:2016npn}. 
The latest multi-lepton searches for
electroweakino~\cite{Sirunyan:2018ubx,Sirunyan:2017lae,Sirunyan:2017qaj,Sirunyan:2017zss} are further applied because of the dominated multi-lepton final states in our model.

In the model, the extra one-loop contributions to muon $g-2$ is given as \cite{mu2h1}
 \beq
    \Delta a_\mu({\rm 1loop}) =
    \frac{1}{2 \pi^2} \, \sum_i
     \kappa_\mu^2 \, r_{\mu}^i \, F_i(r_{\mu}^i),
\label{amuoneloop}
\end{equation}
where $i = H,~ A ,~ H^\pm$, $r_{\mu}^ i =  m_\mu^2/M_i^2$. For
$r_{\mu}^i\ll$ 1 we have
\beq
    F_{H}(r) \simeq- \ln r - 7/6,~~
    F_A (r) \simeq \ln r +11/6, ~~
    F_{H^\pm} (r) \simeq -1/6.
    \label{oneloopintegralsapprox3}\nonumber 
\eeq
The contributions of the two-loop diagrams with a closed fermion loop are given by  
\beq
    \Delta a_\mu({\rm 2loop})
    = \frac{m_{\mu}}{8 \pi^2} \, \frac{\alpha_{\rm em}}{\pi}
    \, \sum_{i,\ell}  Q_\ell^2  \,  \kappa_\mu  \, \frac{\kappa_\ell}{m_\ell} \,  r_{\ell}^i \,  G_i(r_{\ell}^i),
\label{barr-zee}
\end{equation}
where $i = H,~ A$, $\ell=\tau$, and $m_\ell$ and $Q_\ell$ are the mass and
electric charge of the lepton $\ell$ in the loop. The functions $G_i(r)$ are given in Refs.~\cite{mu2h1-1,mu2h2},
\begin{eqnarray}
    && G_{H}(r) = \int_0^1 \! dx \, \frac{2x (1-x)-1}{x(1-x)-r} \ln
    \frac{x(1-x)}{r}, \\
    && G_{A}(r) = \int_0^1 \! dx \, \frac{1}{x(1-x)-r} \ln
   \frac{x(1-x)}{r}.
\end{eqnarray}

We also consider the contributions of the two-loop diagrams with a closed charged Higgs loop,
and find that their contributions are much smaller than the fermion loop. 
The calculations of $\Delta a_e$ are similar to $\Delta a_\mu$, but for the contributions of the two-loop diagrams, 
we include both $\mu$ loop and $\tau$ loop.  

\begin{figure*}[tb]
\includegraphics[width=16.cm]{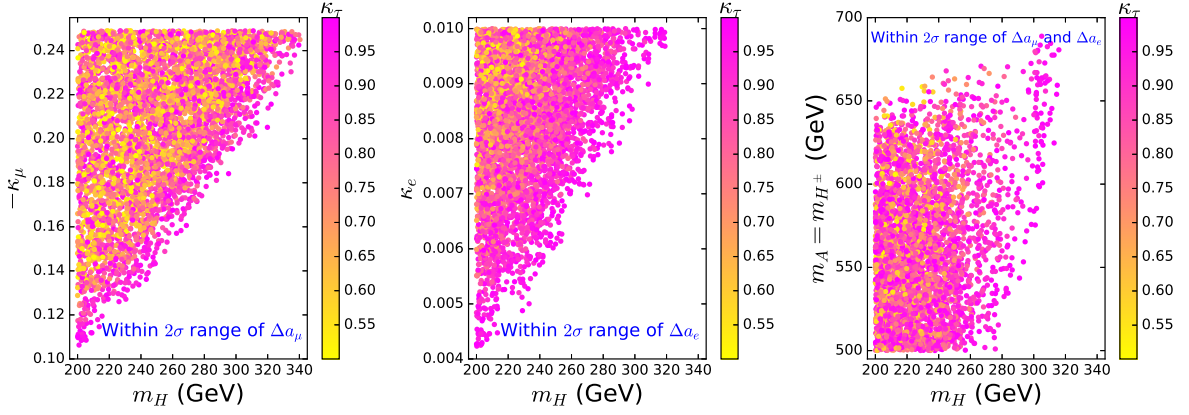}
\vspace{-0.4cm}\caption{The samples within $2\sigma$ ranges of $\Delta a_\mu$ (left panel), $\Delta a_e$ (middle panel), and both
$\Delta a_\mu$ and $\Delta a_e$ (right panel). All the samples satisfy the constraints of the theory and oblique parameters.}
\label{meg2}
\end{figure*}

The HFAG Collaboration reported three ratios from pure leptonic processes, and two ratios
from semi-hadronic processes, $\tau \to \pi/K \nu$ and $\pi/K \to \mu \nu$ \cite{tauexp}. 
In the model, we have the ratios
\begin{eqnarray} \label{deltas-data}
&&\left( g_\tau \over g_\mu \right)^2 \equiv \bar{\Gamma}(\tau\to e
\nu\bar{\nu})/\bar{\Gamma}(\mu\to e \nu\bar{\nu})\approx \frac{1+ 2\delta^\tau_{\rm loop}}{1+ 2\delta^\mu_{\rm loop}}, \nonumber\\
&&\left( g_\tau \over g_e \right)^2  \equiv \bar{\Gamma}(\tau\to \mu
\nu\bar{\nu})/\bar{\Gamma}(\mu\to e \nu\bar{\nu}) \approx \frac{1+ 2\delta_{\rm tree}+ 2\delta^\tau_{\rm loop}}{1+2\delta^\mu_{\rm loop}}, \nonumber\\
&&\left( g_\mu \over g_e \right)^2  \equiv \bar{\Gamma}(\tau\to \mu
\nu\bar{\nu})/\bar{\Gamma}(\tau\to e \nu\bar{\nu}) \approx 1+ 2\delta_{\rm tree}, 
\nonumber\\
&&
\left( g_\tau \over g_\mu \right)_\pi^2= \left( g_\tau \over g_\mu \right)_K^2=\left( g_\tau \over g_\mu \right)^2.
\end{eqnarray} 
Here $\bar{\Gamma}$ denoting the partial width
normalized to its SM value. $\delta_{\rm tree}$ and $\delta_{\rm loop}^{\tau,\mu}$ obtain corrections from
the tree-level and one-loop diagrams mediated by the charged Higgs, respectively. They are given as \cite{tavv-1,tavv-2}
\begin{eqnarray} \label{tree-tau}
\delta_{\rm tree} &=& { v^4 \kappa^2_\tau \kappa^2_\mu \over 8 m^4_{H^\pm}} 
- {v^2 m_\mu \over m^2_{H^\pm} m_\tau} \kappa_\tau \kappa_\mu {g(m_\mu^2/m^2_\tau) \over f(m_\mu^2/m_\tau^2)}, \\
\delta_{\rm loop}^{\tau,\mu} &=& {1 \over 16 \pi^2}  \kappa^2_{\tau,\mu}
\left[1 + {1\over4} \left( H(x_A)  + H(x_H)\right)
\right]\,, \label{loop-tau}
\end{eqnarray}
where $f(x)\equiv 1-8x+8x^3-x^4-12x^2 \ln(x)$, $g(x)\equiv 1+9x-9x^2-x^3+6x(1+x)\ln(x)$, and
$H(x_\phi) \equiv \ln(x_\phi) (1+x_\phi)/(1-x_\phi)$ with $x_\phi=m_\phi^2/m_{H^{\pm}}^2$.

\begin{figure}[h!]
	\centering
	\includegraphics[width=\linewidth]{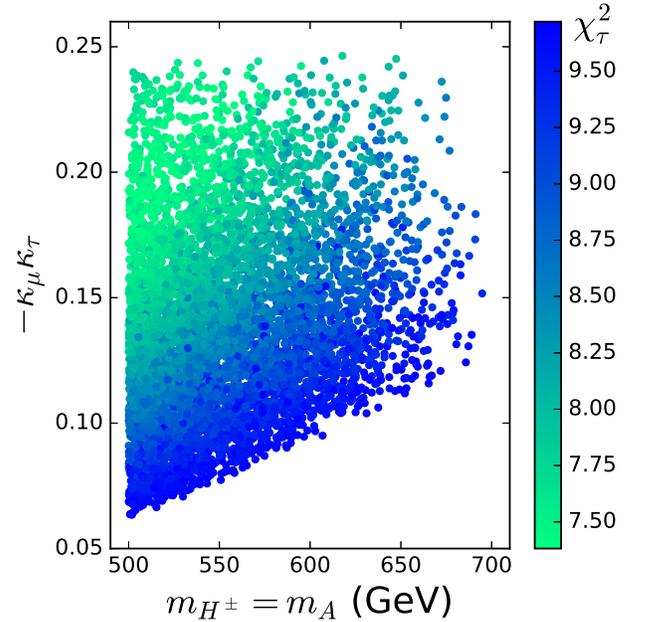}
\vspace*{-0.8cm}	\caption{The surviving samples fit the data of LFU in $\tau$ decay within the $2\sigma$ range. 
All the samples satisfy the constraints of the theory and oblique parameters.}
\label{tauyes}
\end{figure}

\begin{figure*}[tb]
\includegraphics[width=16.cm]{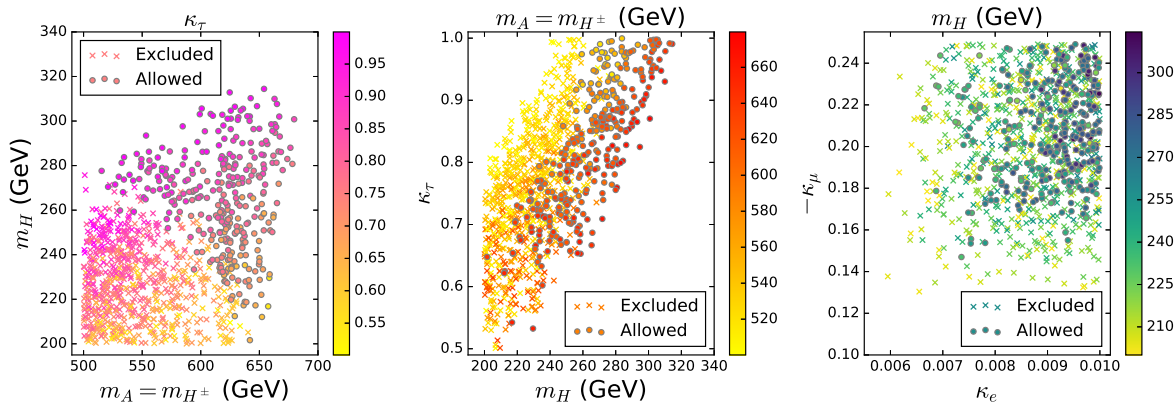}
\vspace{-0.6cm}\caption{The allowed samples (dots with gray edge) and 
excluded samples (crosses) by the direct search limits from the LHC at
95\% confidence level.
The colors indicate $\kappa_{\tau}$, $m_A$ and $m_{H}$ in left, middle, and right panel, respectively.
 All the samples satisfy the constraints of theory, the oblique parameters, $\Delta a_\mu$, $\Delta a_e$, 
the data of LFU in $\tau$ decays, and $Z$ decay.}
\label{zyes}
\end{figure*}

The correlation matrix for the above five observables is 
\begin{equation} \label{hfag-corr}
\left(
\begin{array}{ccccc}
1 & +0.53 & -0.49 & +0.24 & +0.12 \\
+0.53  & 1     &  + 0.48 & +0.26    & +0.10 \\
-0.49  & +0.48  & 1       &   +0.02 & -0.02 \\
+0.24  & +0.26  & +0.02  &     1    &     +0.05 \\
+0.12  & +0.10  & -0.02  &  +0.05  &   1 
\end{array} \right).
\end{equation}
We perform $\chi^2_\tau$ calculations for these five observables. 
The covariance matrix constructed from the data of Eqs.~(\ref{hfag-data})
and (\ref{hfag-corr}) has a vanishing eigenvalue, and the corresponding degree 
of freedom is removed in our calculation.
In our discussions we require $\chi^2_\tau<9.72$, which corresponds to be within
the $2\sigma$ range for four observables, and is smaller than the SM value, $\chi^2_\tau({\rm SM})=12.25$.

The measured values of the ratios of the leptonic $Z$ decay
branching fractions are given as \cite{zexp}
\begin{eqnarray}
{\Gamma_{Z\to \mu^+ \mu^-}\over \Gamma_{Z\to e^+ e^- }} &=& 1.0009 \pm 0.0028
\,,\nonumber\\ 
{\Gamma_{Z\to \tau^+ \tau^- }\over \Gamma_{Z\to e^+ e^- }} &=& 1.0019 \pm 0.0032
\,, \label{lfu-zdecay}
\end{eqnarray}
with a correlation of $+0.63$. 
In the model, the width of $Z\to \tau^+\tau^-$ can have sizable deviation from the SM value
due to the loop contributions of the extra Higgs bosons, because they strongly interact with charged
leptons. The calculations of quantities in Eq. (\ref{lfu-zdecay}) are similar to Ref. \cite{1809.05857}.

After imposing the constraints of the theory and 
the oblique parameters, in Fig. \ref{meg2} we show the surviving samples which are 
consistent with $\Delta a_\mu$ and $\Delta a_e$ at $2\sigma$ level.
Both one-loop and two-loop diagrams  
give positive contributions to $\Delta a_\mu$.
For $\Delta a_e$, the contributions of one-loop are positive and those of two-loop are negative.
Only the contributions of two-loop can make $\Delta a_e$ to be within $2\sigma$ range.
$\Delta a_\mu$ and $\Delta a_e$ respectively favor
negative $\kappa_\mu$ and positive $\kappa_e$ for increasing $m_H$, 
and $m_H$ is required to be smaller than 320 GeV from $\Delta a_e$. 
A large mass splitting between $m_A$ and $m_{H}$ can lead to sizable corrections to $\Delta a_\mu$ and $\Delta a_e$.
Therefore, the right panel of Fig. \ref{meg2} shows that $m_A$ is favored for increasing $m_H$, especially for a large $m_H$.

After imposing the constraints of the theory and 
the oblique parameters, we show the surviving samples with $\chi_\tau^2<$ 9.72 in Fig. \ref{tauyes}.
Such samples fit the data of LFU in $\tau$ decay within $2\sigma$ range. 
Because $\kappa_\mu$ is opposite in sign from $\kappa_\tau$, the second term of $\delta_{\rm tree}$ in Eq. (\ref{tree-tau}) is positive,
which gives a well fit to $g_\tau/g_e$. 
 Fig. \ref{tauyes} shows that $\chi_\tau^2$ can be as low as 7.4, which is much smaller than the SM value (12.25). 
The value of $\chi_\tau^2$ decreases with an increase of $-\kappa_\mu\kappa_\tau$ and increases with $m_{H^\pm}$.

In Fig. \ref{zyes} we show the surviving samples after imposing the constraints of theory, the oblique parameters, 
$\Delta a_\mu$, $\Delta a_e$, 
the data of LFU in $\tau$ decay and $Z$ decay, and the direct searches at LHC.
The model can give sizable corrections to $Z\to \tau^+\tau^-$ for large $\kappa_\tau$ and mass 
splitting between $m_A$ and $m_{H}$. Therefore, the region of the small $m_H$ and large $\kappa_\tau$ is excluded by 
the data of LFU in $Z$ decay, as shown in the middle panel of Fig. \ref{zyes}. The left panel of  Fig. \ref{zyes}
shows that the exclusion limits from the direct searches at LHC favor large $m_H$, $m_A$, and $m_{H^{\pm}}$. 
After imposing the theoretical constraint and relevant experimental constraints,
 the model can explain the anomalies of $\Delta a_\mu$, $\Delta a_e$ and LFU in the $\tau$ decay 
 in many parameter space of 200 GeV $<m_H<$ 320 GeV, 500 GeV $<m_A=m_{H^{\pm}}<680$ GeV, 0.0066 $<\kappa_e<0.01$,  
-0.25 $<\kappa_\mu<-0.147$, and 0.53 $<\kappa_\tau<1.0$. By normalizing event yields in the signal 
regions of Ref. \cite{Sirunyan:2017lae} to higher luminosities, we find that these parameter space can be
fully detected at 95\% confidence level with about 80 fb$^{-1}$ 13 TeV LHC data. 

Note the $Z_2$ breaking term $\mu(\Phi_1^{\dagger} \Phi_2 + \rm h.c.)$ is inevitable 
when we consider the renormalization of one-loop divergent integral.
Although it can be 
set to be zero at some energy scale, radiative corrections will
regenerate it at different scales. 
We can denote it as $\mu$ problem in our model. The vanishing of $\mu$ term does not 
induce an enhanced symmetry so that nothing prevents it to be large via quantum corrections. However, 
we have to figure out that two-Higgs-doublet model is not UV consistent theory since the Higgs mass hierarchy 
problem is not solved. These two hierarchy problem i.e. $\mu$ problem and Higgs mass problem motivate
us to consider new physics around TeV such as supersymmetry, which will be studied in the future paper. 
Therefore, the intrinsic cut-off for quantum 
correction is around TeV. The mixing mass term like "$hH$" can be generated at one-loop by the exchange of SM leptons in the loop, but is 
sizably suppressed by the loop factor of $\frac{1}{16\pi^2}$ and $h\tau\bar{\tau}$ coupling of $\frac{m_\tau}{v}$.
For the cut-off of TeV, we can obtain a small value of $\mu$ 
through the cancellation between bare term and quadratic loop correction. 
The price we paid is that we have to accept fine-tuning. As a result, 
we can still obtain a $Z_2$ symmetric model with $Z_2$ breaking terms being very small.

{\bf Conclusion~--}~We have proposed a lepton-specific inert 2HDM,
where an inert Higgs doublet field with a discrete $Z_2$ symmetry is introduced 
to the SM. Considering all the current theoretical and experimental constraints,
we showed that our model can provide a simple explanation for the anomalies 
of muon $g-2$, electron $g-2$, and LFU of the $\tau$
decays in many viable parameter spaces.

{\bf Acknowledgments~--}~We thank Bin Zhu for help discussions. This work was supported
 by the Projects 11475238,  11575152, 11647601, and 11875062 supported by the 
National Natural Science Foundation of China, 
by the Natural Science Foundation of
Shandong province (ZR2017MA004, ZR2017JL002), by 
the Key Research Program of Frontier Science, CAS,
and  by the ARC Centre of Excellence for Particle Physics at the Tera-scale under the grant CE110001004. 


\end{document}